\documentclass{svjour3}                     
\smartqed  
\usepackage{graphicx}
\newcommand{\be}{\begin{eqnarray}}
\newcommand{\ee}{\end{eqnarray}}

\begin{document}

\title{
Short range correlations
and wave function factorization
in light and finite nuclei 
\thanks{
Presented by S. Scopetta at the 21st European Conference on Few-Body Problems in Physics, Salamanca, Spain, 30 August - 3 September 2010
}}



\author{C.~Ciofi~degli~Atti, L.P. Kaptari, H.~Morita~and~S.~Scopetta}


\institute{  
C. Ciofi degli Atti, L. P. Kaptari, S. Scopetta
\at
Dipartimento di Fisica, Universit\`a degli Studi di Perugia
and INFN, Sezione di Perugia, Italy
\email{ciofi@pg.infn.it; kaptari@pg.infn.it; scopetta@pg.infn.it}           
\and
H. Morita
\at
Sapporo Gakuin University, Japan
\email{hiko@earth.sgu.ac.jp}
}

\date{Received: date / Accepted: date}

\maketitle

\begin{abstract}
Recent BNL and Jlab data provided new evidence on 
two nucleon correlations (2NC) in nuclei.
The data confirm the validity of the convolution model,  
describing the spectral function (SF)
of a correlated pair moving in the mean field 
with high and low relative and center-of-mass (cm) 
momenta, respectively.
The model is built assuming that the wave function (WF) of a nucleus A, 
describing a configuration where the cm momentum of a correlated pair is low
and its relative momentum is high, 
factorizes into the product of the two-body WF and that of the 
A-2 system. Such a factorization has been shown to occur in nuclear matter
(NM).
Here it is shown that few-body systems exhibit
factorization, which seems to be therefore a general property, to be 
reproduced also in studies of the WF of finite nuclei.

\end{abstract}
\vskip 0.5 cm
A new generation of semi-inclusive and exclusive 
electron scattering experiments off nuclei is 
accounting for 2NC with 
unprecedented accuracy (see, e.g., Ref. \cite{sube}).
This activity could have important implications for studies of 
cold dense nuclear matter.
\begin{figure*}
\includegraphics[width=0.5\textwidth]{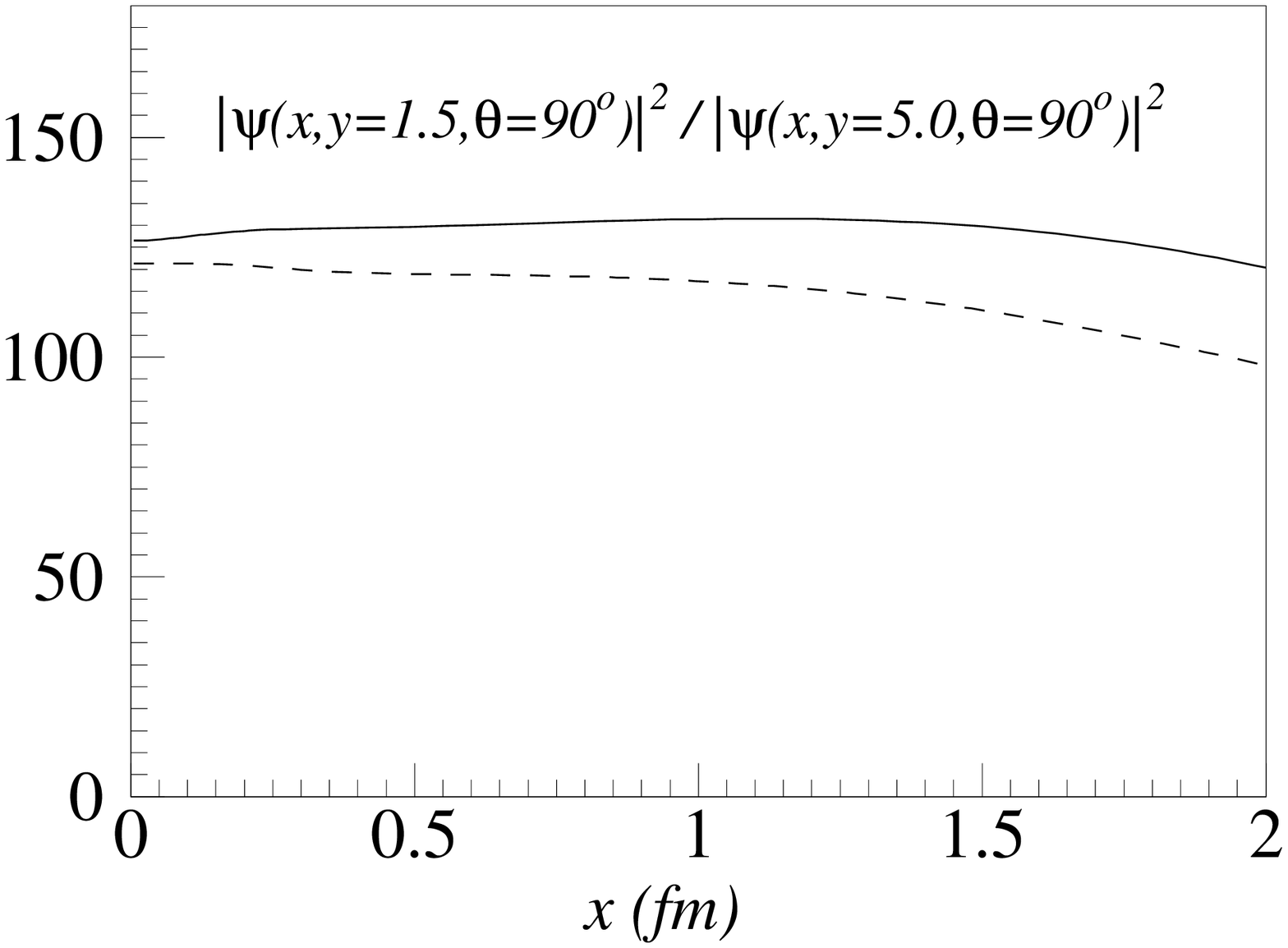}
\includegraphics[width=0.5\textwidth]{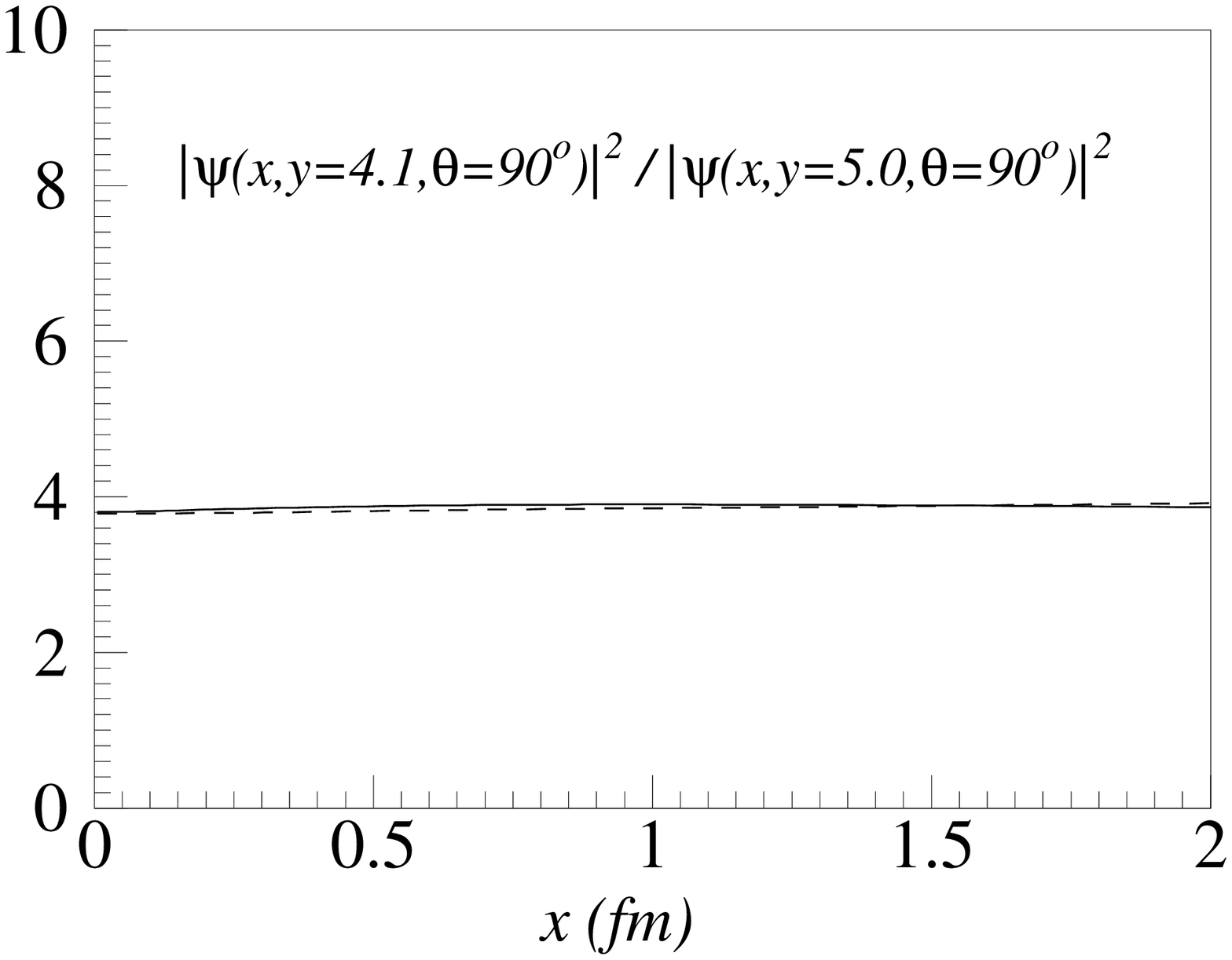}
\vskip -3.5cm
\caption{The ratio Eq. (5) is shown for small values of $x$,
for $\theta_{ \widehat{ {\bf x} {\bf{y}} } }$ = 90 $^o$
and for $y' = 5$ fm.
Left panel: 
$y = 1.5$ fm (far from the 2NC region);
right panel: $y = 4$ fm (in the 2NC region)
dashed line: the same ratio evaluated assuming that the cm motion
is in $S$ wave.
}
\label{fig:1}       
\end{figure*}
The nuclear SF is an important
quantity for studies of 2NC. 
For a nucleon with momentum ${\bf k}$ and removal energy $E$
in a nucleus $A$, 
the SF is defined as follows
\begin{eqnarray}
    {{P(k,E)}} 
&\equiv& {1 \over 2 J_0 + 1} \sum_{M_0 \sigma} ~ \langle \Psi_A^0
    | a_{\vec{k}, \sigma}^{\dagger} ~  \delta[E - (H - E_A)] ~ a_{\vec{k},
    \sigma} | \Psi_A^0 \rangle
\nonumber
 \\
&\equiv& 
{1 \over 2 J_0 + 1} \sum_{M_0 \sigma} ~ \sum_f \left |
    \langle \Psi_{A-1}^f | a_{\vec{k}, \sigma} \right | \Psi_A^0 \rangle |^2
    ~ \delta[E - (E_{A-1}^f - E_A)]  
\end{eqnarray}
{{It is clear from this definition that both the exact wave function (WF) of 
the ground state and the continuum
WF of the $A-1$ system are necessary to evaluate
the SF. These WF are very difficult to calulate, and models 
of the SF can be very helpful.}}
{2NC} are 2-body properties, while
${{P(k,E)}}$ is a 1-body quantity.
Anyway, in the so called 2NC model \cite{pr},
it is argued that,
at high values of $k$ and $E$,
${ {P(k,E)}}$  is
dominated by 
ground-state configurations where  the high momentum
of one nucleun, $\vec{k}_1 \equiv
\vec{k}$, is entirely balanced by that of another one, $\vec{k}_2
\simeq - \vec{k}$, while the remaining ($A-2$) system
has a momentum $\vec{k}_{A-2} \simeq 0$~
\cite{pr}. This yields
\begin{eqnarray} 
   { P(k,E)} 
& \simeq & { P_{2NC}(k,E)} = 
{1 \over 4 \pi} n(k) ~ \delta[E - E^{(2NC)}(k)]
\end{eqnarray}
where $n(k)$ is the nucleon momentum distribution
and energy conservation, $E_{A-1}^* +
{k^2 \over 2 M_{A-1}} = {k^2 \over 2 M}$, fixes
$
E^{(2NC)}(k) = |E_A| - |E_{A-2}|
 ~ + ~ {A - 2 \over A-1} ~ {k^2 \over 2M}~.
$
For these values of $E$,
at high values of $k$,
the exact nuclear 
{SF} of NM and {{$^3$He}}
show indeed a maximum,  
supporting the idea 
of the realization of a 2NC-dominated configuration.
The convolution model of Ref. \cite{cs} 
is a refined version of the 2NC one.
At high values of $k$ and $E$, the two correlated particles 
move in the mean field created by the slow $(A-2)$ system. 
To model the SF, approximations are needed  
for the WF of the ground state and that 
of the  $(A-1)$ system.
As for the ground state WF,
taking the motion of the cm of the pair in S wave 
$\chi_0(\vec{y})$, and
the {$A-2$} system in a low excitation state,
$\Psi_{A-2}^{\bar{0}}( \{
\vec{r}_i \}_{A-2} )$, yields the  
factorized expression
\be
    \Psi_A^0(\{ \vec{r}_i \}_A) \simeq \hat{\cal{A}} \left \{
    \chi_0(\vec{y}) ~ \left [ \Phi(\vec{x}) \otimes \Psi_{A-2}^{\bar{0}}( \{
    \vec{r}_i \}_{A-2} ) \right ] \right \}
\ee
As for the $A-1$ states, 
the interaction between the $A-2$ system
and the correlated particles is neglected.
Using these arguments, the spectral function turns out to be given
by a convolution
\be
    {{P_{FNC}(k,E)}} & = & 
\int d \vec{k}_{cm}
    ~ \delta \left [ E - 
|E_A| + |E_{A-2}|
- {(A-2) \over 2M (A-1)} \left (
    \vec{k} + {(A-1) \vec{K}_{cm} \over (A-2)} 
\right ) ^2 \right ]  \nonumber \\
    & \cdot & 
{{n_{rel} \left (
\left | \vec{k} + {\vec{K}_{cm} \over 2} \right |
\right ) ~
    n_{cm}(| \vec{K}_{cm} |)}}~,
 \ee
where the argument of the $\delta $ function
has been obtained from energy conservation, and
${\bf k}_{rel}=({{\bf k}_1 - {\bf k}_2})/{2}~,   
{\bf K}_{cm}={\bf k}_1 +{\bf k}_2=-{\bf k}_3~$.
One should notice that
the simple $2NC$ model is recovered
placing $n_{cm}(\vec{k}_{cm}) = \delta(\vec{k}_{cm})$, and
that
$P_{FNC}(k,E)$ is
governed by the behaviour of the cm and relative
momentum distributions,
$n_{cm}$ 
and
$n_{rel}$, respectively,
whose
calculation requires the
knowledge of the ground-state wave function only.
One should remember that, to evaluate the exact SF, 
the exact WF of the ground state, as well as the continuum
WF of the $A-1$ system, are necessary.
In the 2NC region, good agreement between $P_{FNC}(k,E)$ 
and the exact calculation for NM and $^3$He is achieved \cite{cs}.
Experimentally, a recent confirmation of the model
has been obtained  at BNL for $^{12}C$ \cite{pia}.
Formally, the model has been justified for NM \cite{bal}.
As for
Few Body Systems, the case of  
$^3$He and $^4$He is discussed here.
More details will be shown in a forthcoming paper
\cite{ckms}.
\begin{figure*}
\includegraphics[width=0.5\textwidth]{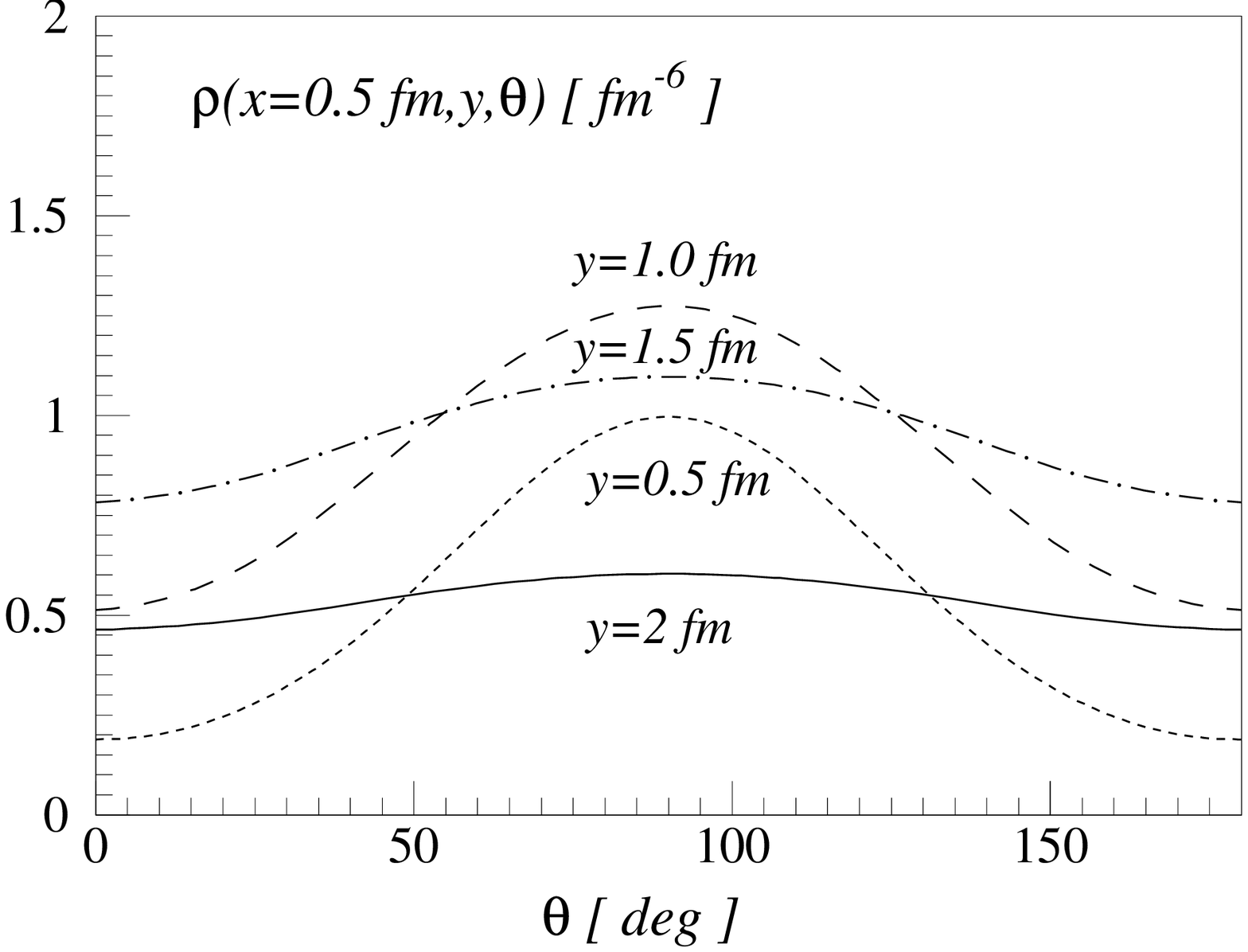}
\vskip -8.5cm
\hskip 6.cm
\includegraphics[width=0.5\textwidth]{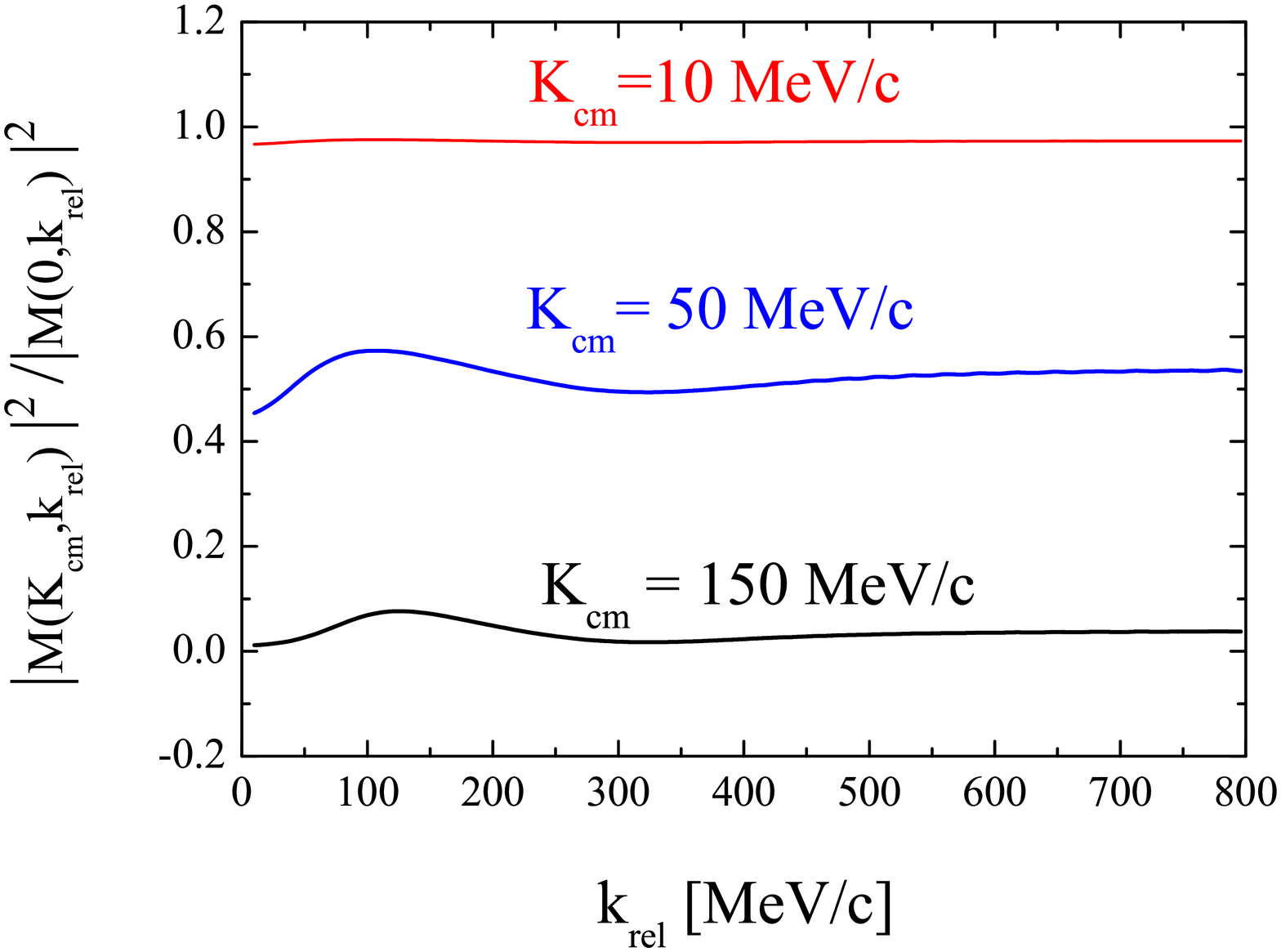}
\caption{Left panel: the quantity
$\rho(x,y,\cos \theta_{ \widehat{ {\bf x} {\bf{y}} } }) $ (see text),
for $x$ = 0.5 fm, for four different values of $y$, as a function of
$\theta_{ \widehat{ {\bf x} {\bf{y}} } } $.
Right panel: the ratio Eq. (9), as a function of $k_{rel}$,
for three different values of ${\bf K}_{cm}$. }
\label{fig:2}       
\end{figure*}
For the convolution model to hold,
in the 2NC region, the WF has to factorize.
For $^3$He, this means that, 
when $x = |\vec r_1 - \vec r_2| << y = |\vec r_1 + \vec r_2|/2$,
$
\Psi_{{\frac12}{{\mathcal M}}}
({\bf x},{\bf y}) \simeq \psi_x({\bf x})\cdot \psi_y ({\bf y})
$
In the following, the occurrence of this property will be studied.
Use will be made of the ``Pisa'' wave function \cite{pisa}, 
corresponding 
to the AV18 interaction \cite{av18}.
If the WF factorizes in the 2NC region,
also the quantity
$\rho(x,y,\cos \theta_{ \widehat{ {\bf x} {\bf{y}} } })
=
\frac{1}{2} \sum_{{\mathcal M}} \left
|\Psi_{{\frac12}{{\mathcal M}}}( {\bf x},{\bf y} ) \right|^2$ does.
This means that
this quantity has to depend weakly upon the angle 
$\theta_{\widehat{{\bf x \bf y}}}$,
and the dependence upon $|{\bf x}|$ and
$|{\bf y}|$ can be separated.
The formal expresion is cumbersom and will be shown elsewhere
\cite{ckms}.
If the factorization holds in the 2NC region, one should have, for
$x << y, y'$,
\be
R_1(x,y,y',\theta_{\widehat{{\bf x}{\bf y}}}) = {  
\rho(x,y,\cos \theta_{\widehat{{\bf x}{\bf y}}})       
\over
\rho(x,y',\cos \theta_{\widehat{{\bf x}{\bf y}}}) 
}
\simeq {f(y) \over f(y')}=constant
\ee
This behavior,
a clear signature of factorization, is found indeed 
around the 2NC region (see Fig. 1).
It is also found that the
cm motion is mainly in S wave (see Fig. 1) and
that the relative motion is mainly
deuteron-like \cite{ckms}.
Moreover, approaching the 2NC region, where $x << y$, the dependence
upon the angle 
$\theta_{\widehat{{\bf x \bf y}}}$ gets weaker and weaker, and the
independence on the relative directions of ${\bf x}$
and ${\bf y}$ is obtained (see Fig.2, left panel).
Let us discuss now the SF.
In the convolution model, the SF
for $^3$He reads
\begin{eqnarray}
    P_{FNC}(k_1,E)  =  
~ \int d {\bf{K}_{cm}}
    ~ \delta 
\left [ E - {\cal{E}}({\bf{k_1}},{ {\bf{K}_{cm}}}) \right ]
{{
n_{rel} \left ( \left | 
{\bf{k_1}} + { {\bf{K}_{cm}} \over 2} \right | \right ) ~
    n_{cm} ( {\bf{K}}_{cm} )}}~,
    \label{g2nc}
\end{eqnarray}
with ${\cal E} = |E_3| + { 1 \over 4M } \left (
    {\bf{k_1}} + 2 {\bf{K}_{cm}} \right ) $,
while the exact expression is found to be
\begin{eqnarray}
P({ k}_1,E)=
~ \int d {\bf{K}_{cm}}
    ~ \delta 
\left [ E - {\cal{E}}({\bf{k_1}},{ {\bf{K}_{cm}}}) \right ]
{{
\frac{1}{2} 
\sum\limits_{{\mathcal S}={\mathcal M},S_{12},\Sigma_{12},\sigma_3} 
\left|\Phi_{\mathcal{S}}({\bf K}_{cm},{\bf k}_{rel}) \right|^2 }}~,
\end{eqnarray}
where $\Phi_{\mathcal S} ({\bf K}_{cm},{\bf k}_{rel}) $ is the
intrinsic $^3$He wave function
in momentum space.
Comparing Eqs. (5) and (6) it is clear that,
in the 2NC region, for the convolution model 
to hold, one has to find
\be
{{
\frac{1}{2} 
\sum\limits_{{\mathcal S}={\mathcal M},S_{12},\Sigma_{12},\sigma_3} 
\left|\Phi_{\mathcal{S}}({\bf K}_{cm},{\bf k}_{rel}) \right|^2 
\simeq n_{cm}(|{\bf K}_{cm}|)\ n_{rel}(|{\bf k}_{rel}|)}}~.
\label{fact}
\ee
The quantity $\left|\Phi_{\mathcal{S}}({\bf K}_{cm},{\bf k}_{rel}) \right|^2$
has been studied. 
Approaching the 2NC region, where
$|{\bf K}_{cm}| <<
|{\bf k}_{rel}|$,
the dependence
upon the angle 
$
\theta_{\widehat{{\bf K_{cm} k}_{rel}}}
$ 
gets weaker and weaker.
This behavior is the one to be studied in forth-coming experiments,
measuring 
{{$n(|{\bf K}_{cm}|)$}} 
and 
{{$n(|{\bf k}_{rel}|)$}}.
Around the 2NC region, where $|{\bf K}_{cm}| <<
|{\bf k}_{rel}|$ and the convolution model holds,
one finds
\be
R=\displaystyle\frac{|\Phi({\bf K}_{cm},{\bf k}_{rel})|^2}
{|\Phi(0,{\bf k}_{rel})|^2}\simeq
\frac{n_{cm}(|{\bf K}_{cm}|)}{n_{cm}(|{\bf K}_{cm}|=0)}=
constant\cdot n_{cm}(|{\bf K}_{cm}|)
\ee
This allows one to see, for any $|{\bf K}_{cm}|$,
at which value of $|{\bf k}_{rel}|$ the convolution model starts
to hold (see Fig. 2, right panel).  
Using this analysis,
$n(|{\bf K}_{cm}|)$ and 
$n(|{\bf k}_{rel}|)$ can be obtained in the 2NC region \cite{ckms}.

As for the factorization of the $^4$He WF, encouraging
preliminary results have been obtained using an
ATMS wave function corresponding to the RSC interaction \cite{atms}.
A result qualitatively similar to that of $^3$He is found.

Summarizing, workable models of the SF,
even if valid only in a peculiar region,
can be very useful for phenomenological studies.
The convolution model is a good approximation to
the exact spectral function in the 2NC region.
For $^3$He, it has been shown that, in the 2NC region, 
the exact wave function exhibits the factorization properties which 
justify the convolution model.
A corresponding analysis is going on for $^4$He, and preliminary 
results are encouraging.
These findings are useful also for
finite nuclei: the factorization  
into cm and relative momentum distributions
seems to be a general property in the 2NC region, 
and many-body calculations
should reproduce it.

%



\end{document}